\documentclass[journal]{IEEEtran}
\usepackage{cite}
\usepackage{textcomp}
\usepackage[T1]{fontenc}
\usepackage{color}

  \usepackage{graphicx, subfigure}
\usepackage[cmex10]{amsmath}
\begin{document}
%
% paper title
% can use linebreaks \\ within to get better formatting as desired
\title{A 2--20 GHz Analog Lag-Correlator for Radio Interferometry}
%
%
% author names and IEEE memberships
% note positions of commas and nonbreaking spaces ( ~ ) LaTeX will not break
% a structure at a ~ so this keeps an author's name from being broken across
% two lines.
% use \thanks{} to gain access to the first footnote area
% a separate \thanks must be used for each paragraph as LaTeX2e's \thanks
% was not built to handle multiple paragraphs
%

\author{C.~M.~Holler,~M.~E.~Jones,~A.~C.~Taylor,~A.~I.~Harris  \IEEEmembership{Member, IEEE},~and~S.~A.~Maas  \IEEEmembership{Fellow, IEEE}

\thanks{Manuscript received October 13, 2011; revised November 30, 2011}
\thanks{\copyright~2011 IEEE. Personal use of this material is permitted. Permission from IEEE must be obtained for all other uses, in any current or future media, including reprinting/republishing this material for advertising or promotional purposes, creating new collective works, for resale or redistribution to servers or lists, or reuse of any copyrighted component of this work in other works.}
\thanks{C. M. Holler is with University of Applied
  Sciences Esslingen, Robert-Bosch-Str. 1, 73037 G\"oppingen, Germany
  (phone: +49-7161-6791269 fax: +49-7161-6792177 e-mail:
  christian.holler@hs-esslingen.de).}%
\thanks{M. E. Jones and A. C. Taylor are with the Subdepartment of
  Astrophysics, Department of Physics, Oxford University, Denys
  Wilkinson Building, Oxford OX1 3RH, United Kingdom (e-mail:
  mike@astro.ox.ac.uk and act@astro.ox.ac.uk).}%
\thanks{A. I. Harris is with the Department of Astronomy, University of Maryland, 
College Park, MD 20742-2421 USA (e-mail: Harris@astro.umd.edu).}%
\thanks{S. A. Maas is with Nonlinear Technologies, Inc, Long Beach, CA, 
USA (e-mail: s.maas@nonlintec.com).}}%

% note the % following the last \IEEEmembership and also \thanks - 
% these prevent an unwanted space from occurring between the last author name
% and the end of the author line. i.e., if you had this:
% 
% \author{....lastname \thanks{...} \thanks{...} }
%                     ^------------^------------^----Do not want these spaces!
%
% a space would be appended to the last name and could cause every name on that
% line to be shifted left slightly. This is one of those "LaTeX things". For
% instance, "\textbf{A} \textbf{B}" will typeset as "A B" not "AB". To get
% "AB" then you have to do: "\textbf{A}\textbf{B}"
% \thanks is no different in this regard, so shield the last } of each \thanks
% that ends a line with a % and do not let a space in before the next \thanks.
% Spaces after \IEEEmembership other than the last one are OK (and needed) as
% you are supposed to have spaces between the names. For what it is worth,
% this is a minor point as most people would not even notice if the said evil
% space somehow managed to creep in.

% The paper headers
\markboth{\copyright~2011 IEEE Transactions on Instrumentation \& Measurement
}%
{Shell \MakeLowercase{\textit{et al.}}: Bare Demo of IEEEtran.cls for Journals}
% The only time the second header will appear is for the odd numbered pages
% after the title page when using the twoside option.
% 
% *** Note that you probably will NOT want to include the author's ***
% *** name in the headers of peer review papers.                   ***
% You can use \ifCLASSOPTIONpeerreview for conditional compilation here if
% you desire.

% If you want to put a publisher's ID mark on the page you can do it like
% this:
%\IEEEpubid{0000--0000/00\$00.00~\copyright~2007 IEEE}
% Remember, if you use this you must call \IEEEpubidadjcol in the second
% column for its text to clear the IEEEpubid mark.

% use for special paper notices
%\IEEEspecialpapernotice{(Invited Paper)}

% make the title area
\maketitle

\begin{abstract}
\boldmath We present the design and testing of a 2--20~GHz continuum band
 analog
  lag correlator with 16 frequency channels for astronomical
  interferometry. The correlator has been designed for future use with a
  prototype single-baseline interferometer operating at 185--275~GHz. The design uses a broadband Wilkinson divider tree with
  integral thin-film resistors implemented on an alumina substrate,
  and custom-made broadband InGaP/GaAs Gilbert Cell multipliers. The
  prototype correlator has been fully bench-tested, together with the
  necessary readout electronics for acquisition of the output
  signals. The results of these measurements show that the response of the
correlator is well behaved over the band. An investigation of the 
noise behaviour also shows that the signal-to-noise of the system is not limited
by the correlator performance.

\end{abstract}
% IEEEtran.cls defaults to using nonbold math in the Abstract.
% This preserves the distinction between vectors and scalars. However,
% if the journal you are submitting to favors bold math in the abstract,
% then you can use LaTeX's standard command \boldmath at the very start
% of the abstract to achieve this. Many IEEE journals frown on math
% in the abstract anyway.

% Note that keywords are not normally used for peerreview papers.

% For peer review papers, you can put extra information on the cover
% page as needed:
% \ifCLASSOPTIONpeerreview
% \begin{center} \bfseries EDICS Category: 3-BBND \end{center}
% \fi
%
% For peerreview papers, this IEEEtran command inserts a page break and
% creates the second title. It will be ignored for other modes.
\IEEEpeerreviewmaketitle

\section{Introduction}
% The very first letter is a 2 line initial drop letter followed
% by the rest of the first word in caps.
% 
% form to use if the first word consists of a single letter:
% \IEEEPARstart{A}{demo} file is ....
% 
% form to use if you need the single drop letter followed by
% normal text (unknown if ever used by IEEE):
% \IEEEPARstart{A}{}demo file is ....
% 
% Some journals put the first two words in caps:
% \IEEEPARstart{T}{his demo} file is ....
% 
% Here we have the typical use of a "T" for an initial drop letter
% and "HIS" in caps to complete the first word.  

\IEEEPARstart{T}{he cross-correlator} is the heart of an
interferometric radio telescope, where the signals from each antenna
are combined to form the complex visibility measurements from which
the image of the sky is derived.  Interferometry, with its inherently
high stability and direct measurement of spatial spectra, is a
particularly valuable technique for studying fine-scale structure in
the Cosmic Microwave Background (CMB).  By measuring the power
spectrum of the CMB, we can obtain a unique snapshot of the
distribution and motions of matter in the Universe 400,000 years after
the Big Bang, providing constraints on key cosmological
parameters. Results from such observations have helped establish the
overall geometry of the Universe, and therefore our understanding of
its fate; new observations will test models of the earliest moments in
the formation of the Universe.

\begin{figure}[h!]
\begin{center}
\includegraphics[width=3in]{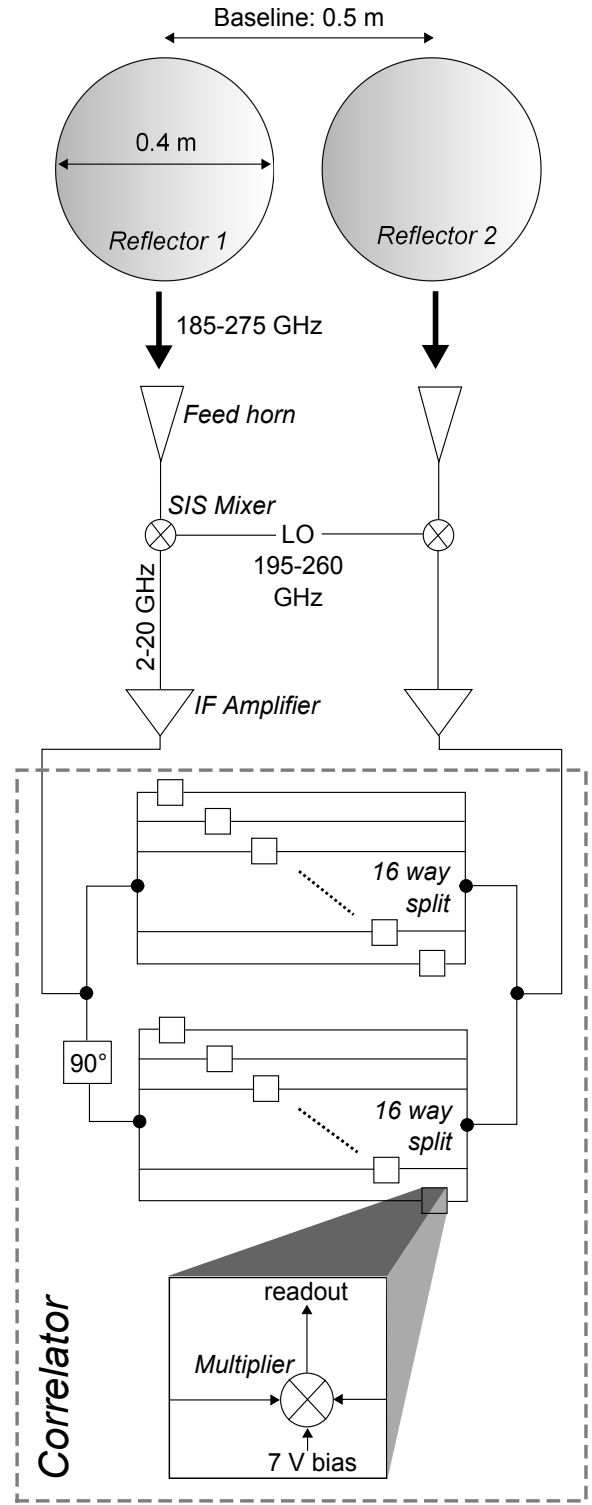}
\end{center}
\caption{
A schematic of the GUBBINS single baseline interferometer including the 16 lag complex Fourier Transform correlator.
}
\label{fig:schematic}
\end{figure}

Improvements in front-end receiving systems and in the back-end signal
processing systems are needed to help progress the study of the
CMB. Increasing the system bandwidth allows more precise radiometry
and hence more sensitive measurements. Here we describe a
cross-correlator that covers the very wide bandwidths needed to gather
signals for measurements of tiny deviations from the CMB’s 2.73~K
blackbody radiation temperature. Wide bandwidths are needed to fully
exploit the windows in the atmospheric transmission in the millimeter
wavebands, which are typically many tens of GHz wide. This correlator
operates over the frequency range 2--20~GHz, and is the back-end for a
prototype wide-band interferometer, GUBBINS \cite{GUBBINS}, 
currently under development. This telescope is a
millimeter-wave (185--275~GHz) single-baseline interferometer
employing SIS mixers and very wide bandwidth low-noise IF amplifiers.
A schematic of the instrument is shown in Fig. \ref{fig:schematic}.

Other broadband analog correlators which have been used in experiments devoted to
the CMB have bandwidths up to 16~GHz
(e.g.  \cite{CBI}, \cite{AMIBA}). The correlator we discuss here has a
combination of broader bandwidth, lower complexity and lower power
dissipation than previous correlators used in CMB interferometry,
properties that enable the construction of larger interferometric
arrays.

\section{Digital vs. Analog Correlators}

Correlators can be implemented using either analog or digital
technology. Analog correlators use voltage multiplier circuits and
integrators to correlate the two signals, while in digital correlators
the signals are digitized, and multiplied and integrated using digital
circuits. 
If the correlated signal must be split into many frequency channels for high resolution spectroscopy, or the path lengths from antennas to the correlator need to be equalized by relatively large time delays, digital correlators are the common choice because of the straightforward hardware requirements. However, analog correlators can offer a more economic solution if there is no requirement for fine frequency resolution or large delays, but
very large continuum bandwidth, low power consumption
and reduced complexity are required.

For example, we
can compare our analog implementation of the single-baseline GUBBINS
correlator with an equivalent implmentation using CASPER\cite{casper},
a hardware/firmware system of digital processing equipment optimized
for radio astronomy, which is currently being used for several large
correlator projects\cite{parsons}. A typical CASPER-based
implementation would employ $1\cdot10^9$ samples/sec, 8-bit analog-to-digital
converters, each of which can sample a 500-MHz baseband signal. To
cover our 2--20~GHz band for two antennas would require 72 such
converters, plus a large amount of RF electronics (mixers, LO combs,
filters) to split and down-convert the signal bands in to appropriate
baseband channels. The signal correlation would be carried out using
field-programmable gate arrays (FPGAs), for example the Xilinx Virtex
5 on the ROACH processing board. Each ROACH can host two dual-channel
A/D converter cards, so 18 ROACHes would be required. Each ROACH board can
provide up to $400\cdot10^9$ op/sec of processing (with full utilization of the
FPGA multipliers). The processing rate required for our correlator is
given by the product of the Nyquist data rate ($36\cdot10^9$ samples/sec), the
number of lag channels (16) and the number of real multiplies per
complex multiply (4), giving $2.3\cdot10^{12}$ op/sec. With realistic utilization
levels this could be comfortably handled by the 18 ROACHes, provided
they were interconnected with an 18-port 10-gigabit ethernet
switch. The total cost of this hardware would exceed \$100,000. 

It can
be seen however, that the bottleneck is more in the signal acquisition
rather than processing; development of low bitcount digitizers with
sampling rates of many tens of gigasamples per second, along with
increases in the rate of digital operations on a single device, may
eventually swing the economics in favour of a digital solution.

\begin{figure*}[t!]
    \begin{center}
        \subfigure[]{%
            \includegraphics[width=3in]{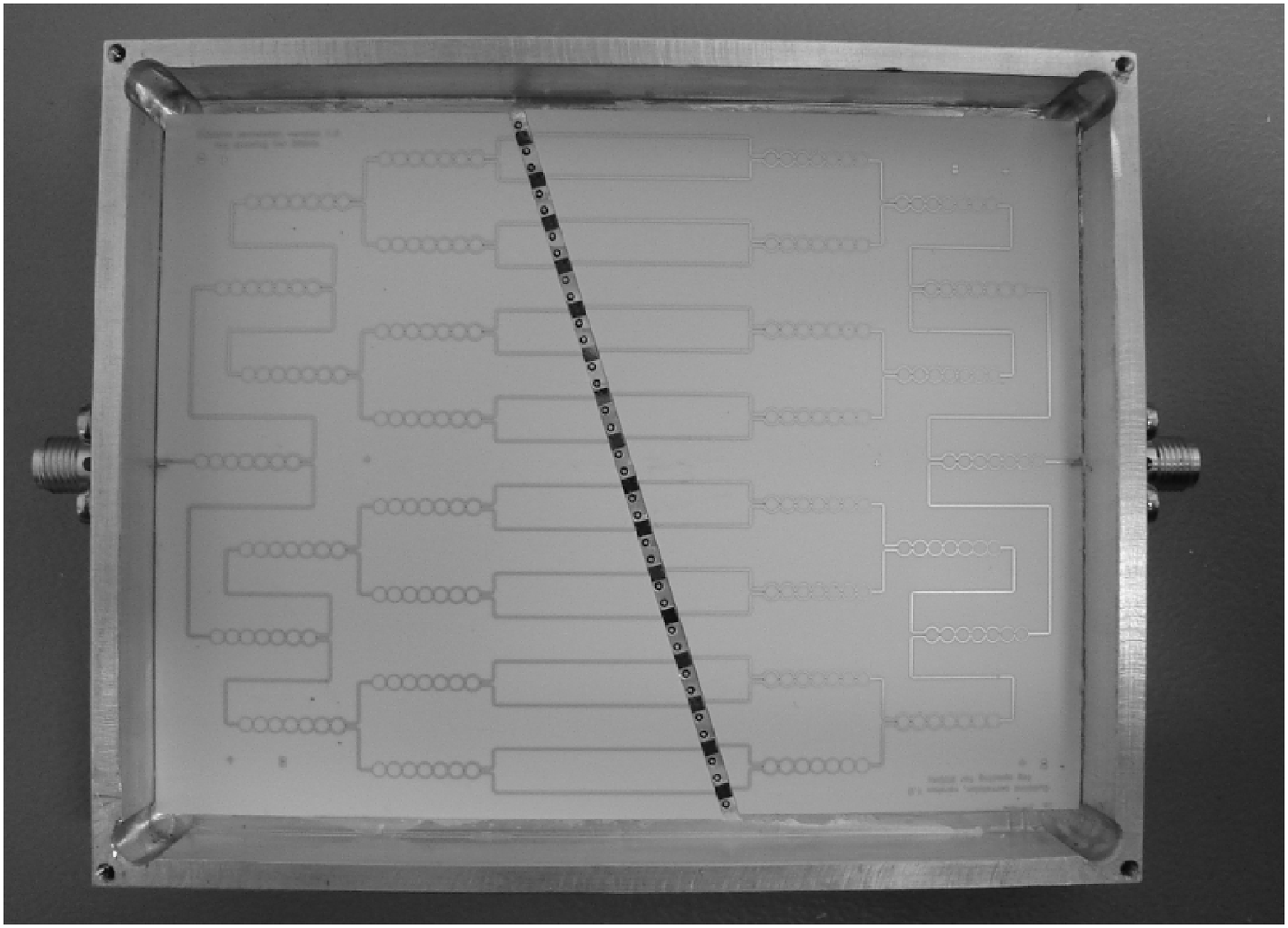}
        }%
        \subfigure[]{%
           \includegraphics[width=3in]{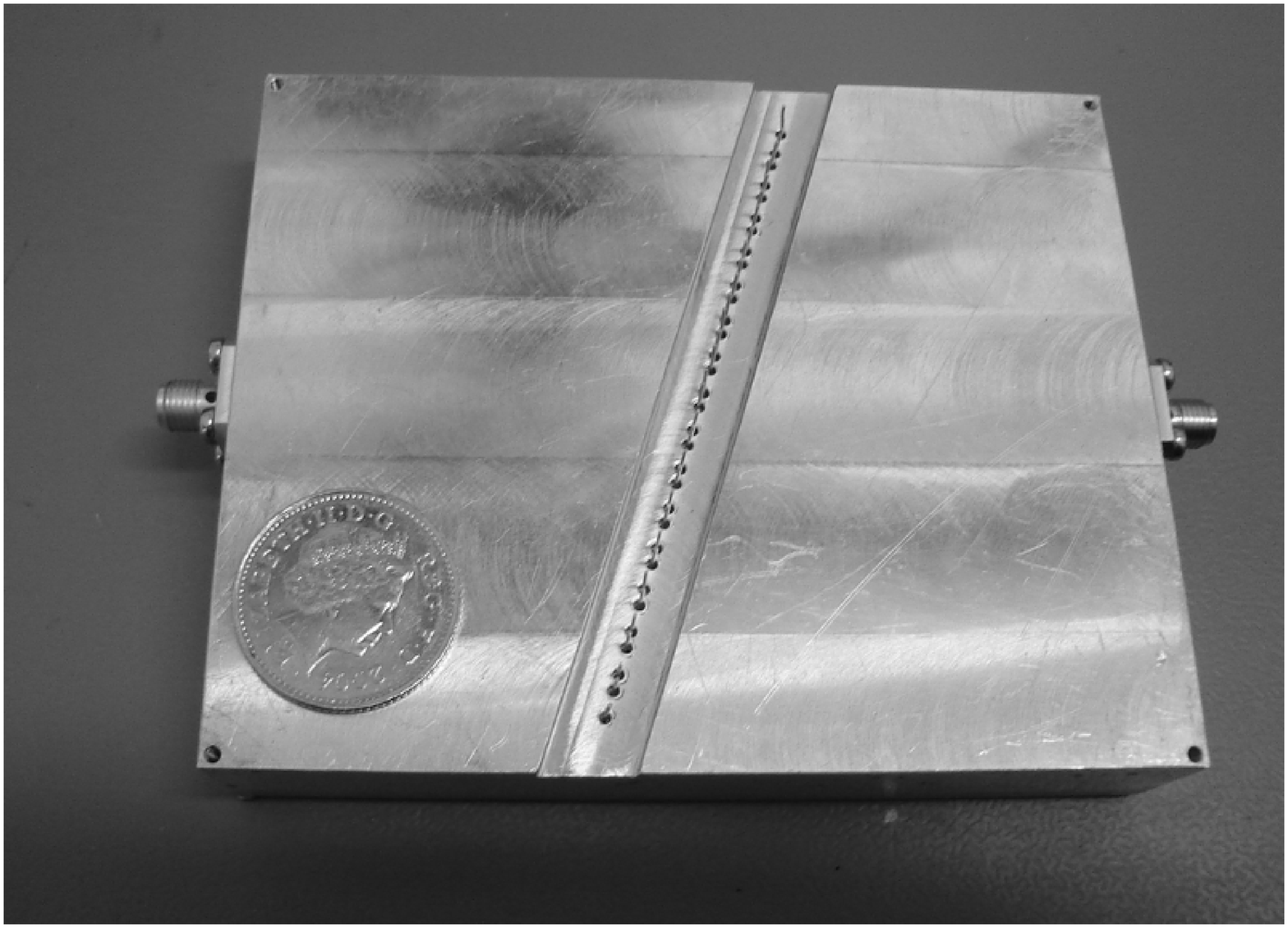}
        }\\ %  ------- End of the first row ----------------------%
        \subfigure[]{%
            \includegraphics[width=3in]{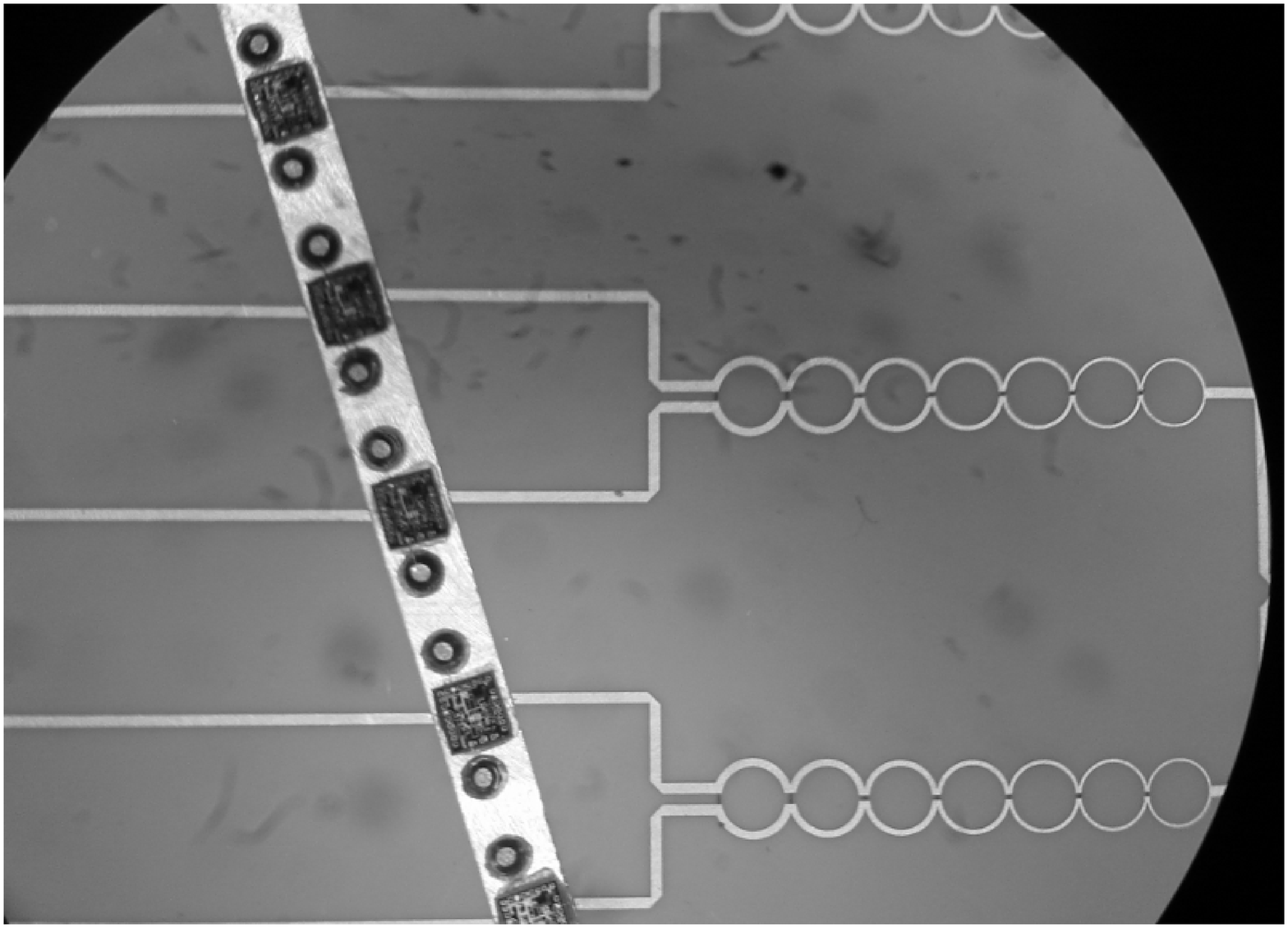}
        }%
        \subfigure[]{%
            \includegraphics[width=3in]{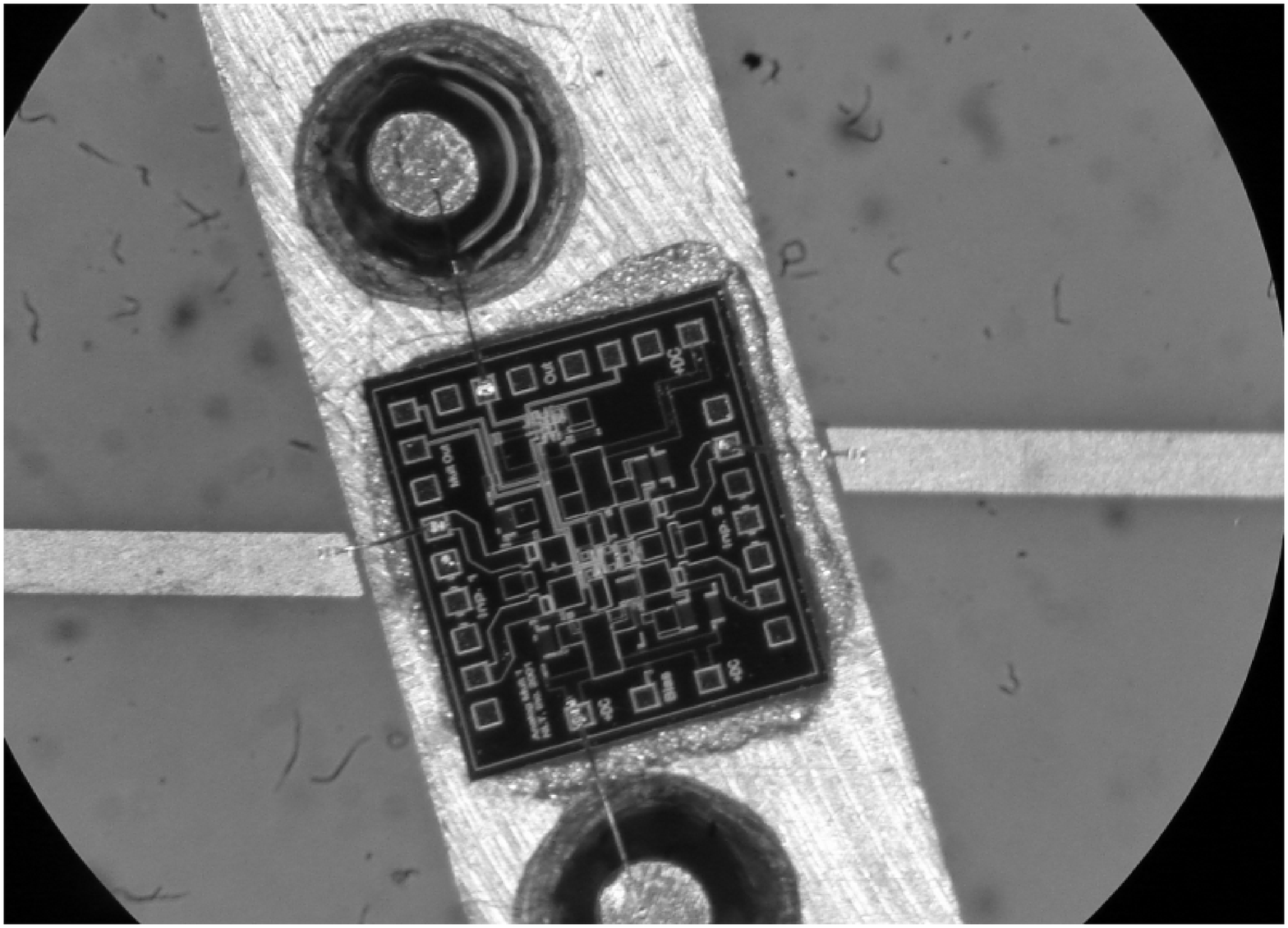}
        }%
    \end{center}
    \caption{%
       (a),(b): photos of a single correlator module showing the Wilkinson splitter tree, multipliers and feed through pins for signal output and bias input on the back side of the box (the coin for scale is 25~mm diameter). (c): high resolution photos of the central part of the correlator including the signal splitter. (d): detailed photo of a multiplier chips showing the bonding to the microstrip lines and feed through pins.
     }%
\label{fig:photo}
\end{figure*}

\section{Correlator Architecture} 

The key design decisions in the implementation of a very broadband
correlator relate to the choice of frequency band, whether the signal
band is divided into narrower channels before or after correlation,
the type of non-linear element used to perform the multiplication, and
the method of physical distribution of the signals to form the
required cross-correlations.

Some previous correlators have achieved high bandwidth by splitting
the IF band in to multiple sub-bands which are downconverted to a
common baseband. For example, the CBI/DASI correlator \cite{CBI}
splits a 2--12~GHz IF band into ten 1--2~GHz bands using
filter-mixers. The signals are then correlated using ten identical
units in which Gilbert cell multipliers are placed at the
intersections of crossing striplines. This method has the advantage of
keeping the bandwidth of the individual correlation units small, but
the filter-mixer bank is a significant additional complication. Given
the recent availability of off-the-shelf microwave components up to
20~GHz, one of our goals was to process a large IF bandwidth without
further explicit splitting and downconversion.

The non-linear multiplying element is also a key design choice. Active
multipliers (Gilbert cells) \cite{CBI}, balanced mixers \cite{AMIBA},
and diodes \cite{AMI} are all possible choices. For this project we
use a 20-GHz bandwidth custom-made Gilbert cell multiplier developed
specifically for use in astronomical correlators. Individual
monolithic microwave integrated circuit (MMIC) chips contain a
four-quadrant Gilbert cell multiplier \cite{Gilbert} fabricated in
InGaP/GaAs heterojunction bipolar transistors (HBT)
technology. Transistor speed limits the upper end of the multiplier's
usable response to between 20 and 25~GHz. The frequency response of
test amplifiers on the same wafers suggested transistor $f_{\rm{max}}$
of about 70~GHz, in reasonable agreement with the foundry's \cite{GCS}
nominal values for their process at the time of the
fabrication in 2002. Integrated bias blocking capacitors at the inputs set the
low-frequency response end to about 1~GHz.  Resistive matching at the
chip edges provides an excellent match across the input band. An
integrated differential postamplifier buffers the multiplier output
and provides further gain.  More detailed information is available in
references \cite{Harris1},\cite{Harris2} and \cite{Harris3}.

Having chosen to use a single wide IF band we need to provide
frequency resolution by means of a lag correlator, in which the
outputs of multiple correlations with different lag spacings are
Fourier transformed to give the cross-frequency spectrum. In general a
lag correlator samples a correlated signal from two antennas at fixed
time delays. These are normally set to the Nyquist rate
$1/(2\nu_{\rm{IF}})$ in order to recover the spectrum of a signal with
bandwidth $\nu_{\rm{IF}}$. However, depending on the type of
interferometer and other hardware restrictions, different delay
structures are possible \cite{TMS}. In general, the relationship
between the interferometer antenna diameter $d$, its longest baseline
$D$, and the RF and IF frequencies dictates the need for negative and
positive, or only positive delay lines. If

\begin{equation}
\frac{D}{d\cdot \nu_{\rm{RF}}} \ll \frac{1}{2\cdot \nu_{\rm{IF}}}, 
\end{equation}

then the additional time delay for a signal originating from the edge
of the beam instead of the center is much smaller than a single lag
delay in the correlator. In this case it is sufficient to sample
positive delays only at Nyquist rate, since the centre of the cross
correlation function (CCF), the so-called zero lag, is fixed in lag
space. All the complex spectral information is contained in the
positive half of the CCF. If the inequality (1) does not hold,
negative and positive delays have to be used in order not to lose
signal at the beam edges, since sources originating at different
places in the beam can introduce positive or negative lags that are
significant compared to the coherence time. In this case the lags have
to be sampled with twice the Nyquist lag spacing.

Both architectures depend on the availability of a broadband
90$^\circ$ phase shifter: the \textit{complex} correlator produces
0$^\circ$ and 90$^\circ$ correlations (Fig. \ref{fig:schematic}),
which correspond to real and imaginary values, respectively. However,
a \textit{real} correlator can be built without the need of a
broadband phase shift, using twice the number of 0$^\circ$
correlations with half the lag spacing and no 90$^\circ$ correlations
\cite{AMI}. Fourier transformation of the measured values results in
the same complex power spectrum.

Experience shows that the use of a complex correlator has practical
advantages, especially in the commissioning stages of a telescope,
since the complex visibility is available before the signal has been
Fourier transformed, and a true phase can thus be measured in each lag
channel separately. For this reason we decided to build a
\textit{complex} lag correlator, using commercially available
90$^\circ$ hybrids, which although relatively expensive are only
required one per antenna. Since equation (1) is valid for the GUBBINS
telescope, i.e. the baselines are short, we only measure positive
lags, using the Nyquist sampling rate.

In the correlator described below, each frequency band is split into
16 complex channels, and so 32 correlations are necessary for a single
baseline. All the correlations for each complex component are combined
on a single circuit board and two such modules per baseline are
required.

The correct time delay of each lag is essential for the frequency
bandwidth of the correlator. The number of lags defines the resolution
of a frequency channel. The sampling rate of our correlator is set to
20~GHz, therefore the channels covering a total frequency range
0--20~GHz have a width of 1.25~GHz each.  In order to suppress
aliasing at the upper frequency limit a steep cut-off at 20 GHz is
necessary. Alternatively the highest one or two channels can be
discarded. For GUBBINS we implemented a filter for signals above
18~GHz. For further details on analog lag correlators see for example
\cite{Harris_WASP} or \cite{AMI}.

Since the response of the multiplier chips only drops off slowly below 2~GHz and above 20~GHz, a somewhat higher bandwidth of approximately 1--23~GHz with the appropriate lag spacing and calibration should in principle be possible, but has not been investigated yet.

\section{Correlator Design}

Photos of a single correlator module (half of a complex correlator) can be seen in Fig. \ref{fig:photo}. On each board two splitter trees of broadband Wilkinson dividers split the incoming signals from the two antennas of a baseline into 16 identical signals. All signals are then correlated after appropriate time delays using the custom-made Gilbert cell multipliers.

\begin{figure}[h!]
\begin{center}
\includegraphics[width=3.2in]{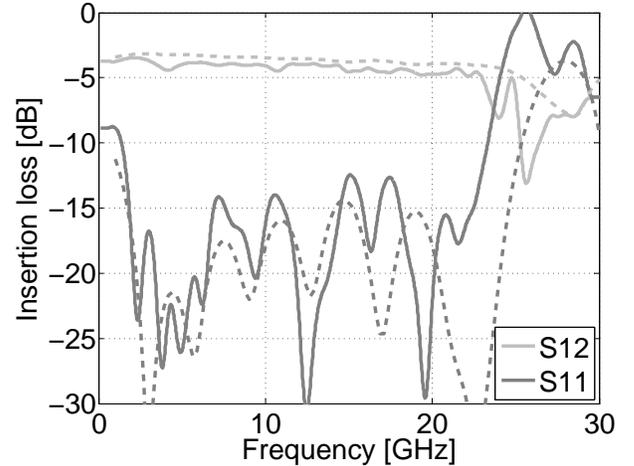}
\end{center}
\caption{
Measurement (solid) and simulation (dashed) of a single seven stage Wilkinson power divider as used in the correlator splitter tree. The difference between measurement and simulation above 20~GHz can be mainly attributed to the connectors.
}
\label{fig:wilkinson}
\end{figure}

The correlator module shown consists of a single-layer microstrip board on a 10-mil alumina substrate with a dielectric constant of $\epsilon_r=9.9$ . The two splitter trees contain fifteen Wilkinson power dividers, arranged in a binary tree. The power dividers use seven elements in order to give sufficient bandwidth with adequate insertion loss and matching, and work up to approximately 23~GHz. Lumped element resistors have too large parasitic reactances to work well over this bandwidth so we use thin film resistive sheet to form the resistive elements between the rings which provide for high output isolation. We experimented with Duroid substrates, using OhmegaPly nickel-phosphorus film underneath the copper traces to form the resistive elements by a pure etching process, but we found that the resulting slope of insertion loss with frequency was excessive. Instead we have adopted alumina substrate with deposited tantalum nitride thin-film resistive elements, which gives much better frequency response. Measured results compared to simulations of a single splitter can be seen in Fig.~\ref{fig:wilkinson}.

Placed between the two splitter trees are 16 multiplier chips (Fig. \ref{fig:schematic} and \ref{fig:photo}). Due to the delay steps they are oriented diagonally to the delay lines. The Nyquist rate here is ${(2\cdot20~\rm{GHz})^{-1}}$  and corresponds to 7.5~mm delay difference from one multiplier to the next (the lag spacing) in vacuum. Due to the high dielectric constant this is reduced to 2.92~mm on the alumina board. Since the signals are counter-propagating, the horizontal distance between two adjacent multiplier chips is only half that distance i.e. 1.46~mm. The signals from the microstrip lines are directly wire bonded to the input pads of the multiplier chips which have $50\Omega$ input impedance. There are two metal feed-through pins on either side of each chip. One is used for the bias supply to the multiplier and the other for the low frequency output signal from the multiplier. These connect to a FR4 circuit board, which is mounted on the back of the correlator box, containing further analog gain, buffering and filtering. The signals are then passed via a backplane to a 16-channel digitizer board where the signals are sampled at 14-bits resolution and 2.8~Msamples/sec, and integrated in an FPGA.

\section{Signals and Phase Switching}

The desired correlated signal of the observed astronomical object is in general deeply buried inside uncorrelated noise.  The undisturbed extraction of this average correlated signal voltage is the goal of a cross-correlator. Uncorrelated noise in the system has different sources. The main contributions are receiver noise, atmospheric temperature, antenna spillover and ohmic losses in the system.

However, there are many additional sources of signal contamination in the signal path after the antenna, which can be strongly correlated, e.g. pick-up of terrestrial signals, temperature fluctuations, mains pick-up and cross talk between antennas. Therefore a phase-switching system is normally introduced at the earliest possible point in the receiver chain in order for these signals to be eliminated. For this the signals from the different antennas are modulated by orthogonal periodic functions, e.g. Walsh functions. 

Phase-switching only suppresses signals with longer coherence time then the repeat period of the switching functions. Therefore, a high phase-switching frequency is preferred. However, for accurate demodulation of the switched signal, the power in the modulated signal must be contained within the video bandwidth of the detectors. If the loss of signal due to filtering is to be less than 0.1\%, a basic frequency of about a thousand times lower than the video bandwidth is required. 

The multiplier's video bandwidth is approximately 10~MHz, but in order to reduce the required sampling frequency of the digitizers (which is fixed at 2.8~Msamples/sec for the sampler board available to us) this is filtered down to 800~kHz and therefore a phase switching frequency of less than 1~kHz is required. Further details in the next section about low frequency noise show that this is very well possible. A more detailed discussion of suppression levels that can be obtained by phase switching can be found in \cite{CBI}.

\section{RF Measurements}

Here we present the results of testing four prototype correlator modules, of which one is shown in Fig. \ref{fig:photo}.

\subsection{Return Loss}

\begin{figure}[h!]
\begin{center}
\includegraphics[width=3.0in]{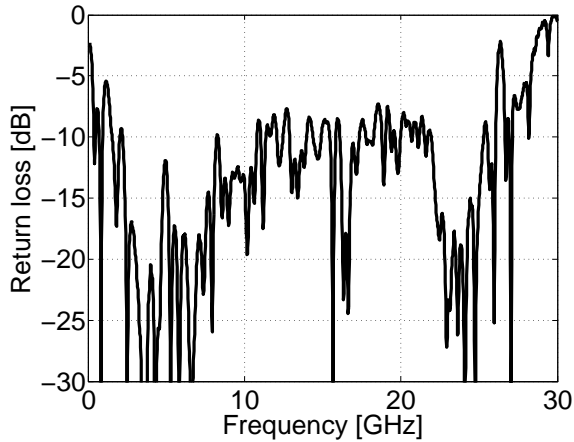}
\end{center}
\caption{Return loss at the correlator input.}
\label{fig:returnloss}
\end{figure}

The return loss at the inputs to the correlator board is shown in Fig. \ref{fig:returnloss}. It is well matched across the band of 2--20~GHz.

\subsection{Lag Spacing}

Regular lag spacing (the electrical path distance between to adjacent multipliers)
is important for the
performance of the correlator. Variations in the lag spacing result in an irregular
Fourier transform, which is not in general inferior to a regular one.
However, in the presence of noise, this leads to a reduction in
signal-to-noise; some power will be scattered from one frequency
channel to another \cite{Harris_WASP}. Calibration schemes 
can minimize this effect of irregular lag spacing.

As mentioned above, delays between the multipliers should ideally be
7.5~mm in free space, or $c/(2\cdot20 \rm{GHz})$, which corresponds to
about 2.92~mm in alumina.  Some degree of distance variation is expected
due to, for example, small irregularities in the printing process of
the microstrip lines, placing of the multiplier chips and variations
in both the dielectric constant of the substrate and the bond length
to the multiplier chips. Dispersion in the microstripline transmission
lines causes the distance also to vary across the band. 

We measure the lags directly by feeding a CW
signal into both inputs to the cross-correlator, with an additional
delay at one input. When the signal is stepped in frequency the
response of a single lag is a sine wave as a function of frequency and
the wavelength of this sine wave depends on the lag’s absolute
distance from the zero lag. The difference in wavelength for different
lags makes it possible to calculate the relative distances very
accurately. For more details see
\cite{Holler_PhD}. 

\begin{figure}[h!]
\begin{center}
\includegraphics[width=3.0in]{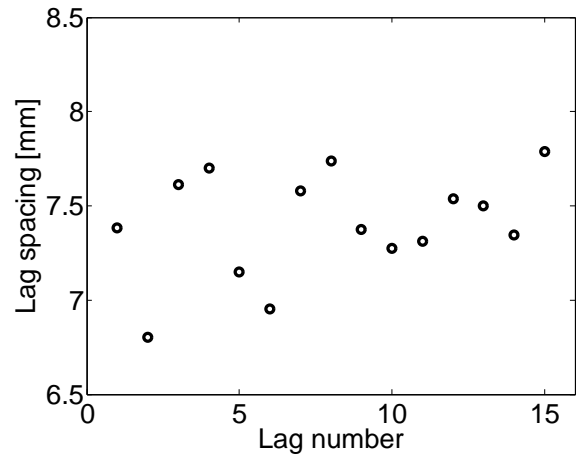}
\end{center}
\caption{Variations in measured lag spacing, the distances between
two adjacent multipliers, for one prototype
  correlator board. The ideal distance for a
  20~GHz bandwidth of the correlator corresponds to 7.5~mm in vacuum.}
\label{fig:lags}
\end{figure}

Fig. \ref{fig:lags} shows the measured delays
between adjacent points of correlation. The average
delay step is 7.40~mm which means that the cut-off frequency of the
Fourier transform is 20.3~GHz. The biggest relative lag error is 9\%
and the standard deviation is 3.8\%. The lag errors are small and
stable with time, which means their effect on the final transformed
spectrum can be removed with suitable calibration. Path and phase
errors due to the external hybrid and fixed cables are also stable and
are removed by the same calibration procedure.

The lag errors are within the expected range and the effect on the recovered signal, namely signal scattering between frequency channels, can be minimized using the correct calibration. Calibration can either employ a set of known frequencies which are used to measure the position (phase) and frequency response (amplitude) of each multiplier, as in these tests, or alternatively a known broadband signal (e.g. an astronomical source) that is swept in delay can be used to measure the same properties. Provided that the lag errors are small compared to the lag spacing, the loss in signal-to-noise in this calibration procedure is small. If the error approaches a whole lag spacing then the information sampled by each lag becomes degenerate and signal-to-noise is permanently lost. 

\subsection{Frequency Response}

\begin{figure}[h!]
\begin{center}
\includegraphics[width=3.05in]{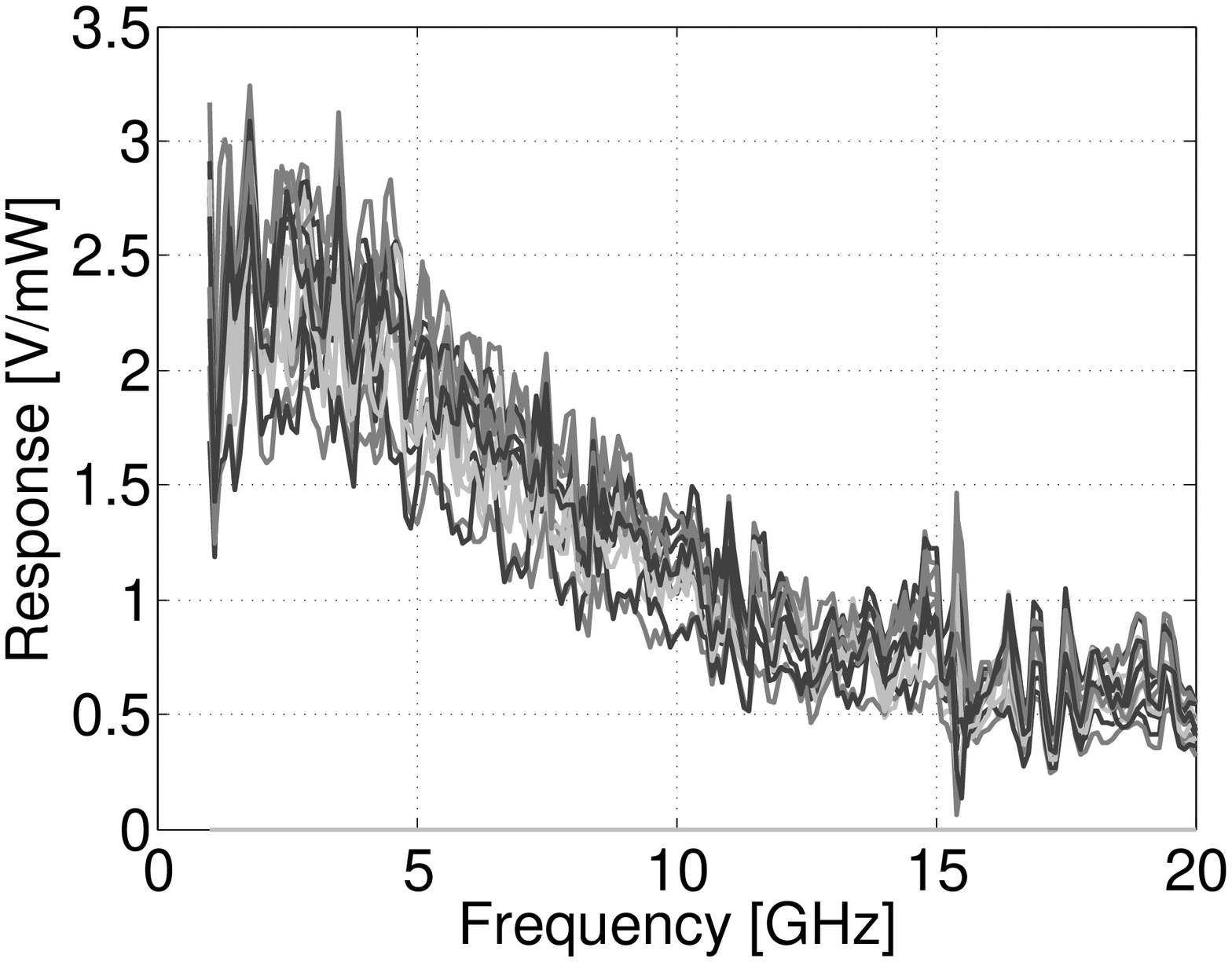}
\\
\includegraphics[width=3.05in]{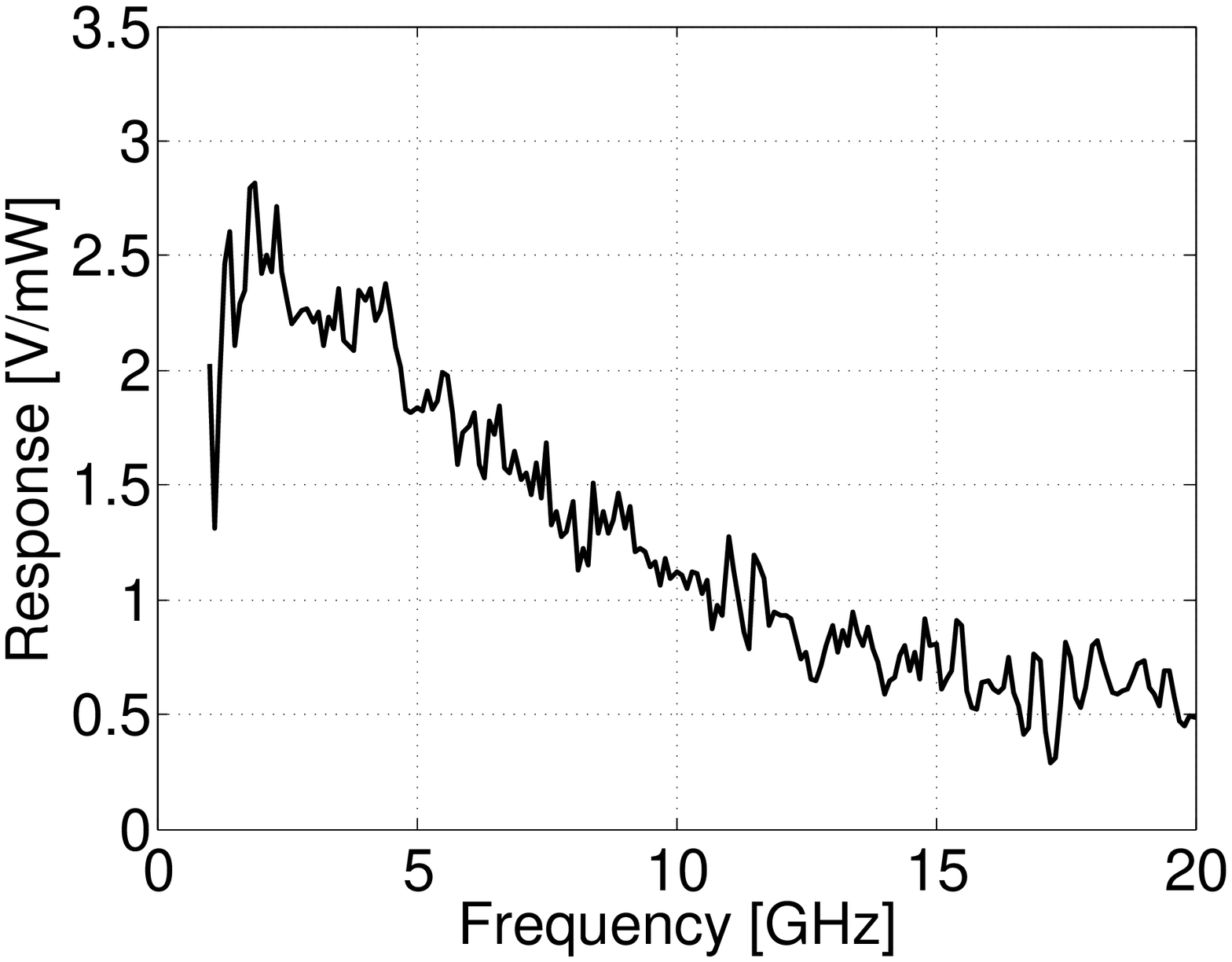}
\end{center}
\caption{Frequency response of all 16 channels of one prototype board (top) and one typical channel (bottom).}
\label{fig:frequ}
\end{figure}

The frequency response of the correlator can be seen in Fig.~\ref{fig:frequ}. The variation in response between the channels is less than 2~dB. Comparing the correlator passband to the response of an individual multiplier shows that the splitter tree contributes approximately $-4$~dB to the slope across the frequency band, the multiplier itself about $-2$ to $-3$~dB. 

There is a narrow resonance at about 15.5~GHz in several of the channels, whose origin is unclear. This resonance could also be seen in the response of a separate single multiplier chip and therefore is not connected to the splitter tree. The resonance is stronger for lower input power. The common structure between 16 and 18~GHz is also present in the separate individual multiplier chips.

The other small scale variations in the passband are caused in part by measurement noise and amplitude and phase errors in the 90$^\circ$ hybrid, which has been used to measure real and imaginary values. Also, some of these variations can be attributed to passband variations of the microstrip circuit, but in general the response is well behaved.

\subsection{Power Response}

\begin{figure}[h!]
\begin{center}
\includegraphics[width=3in]{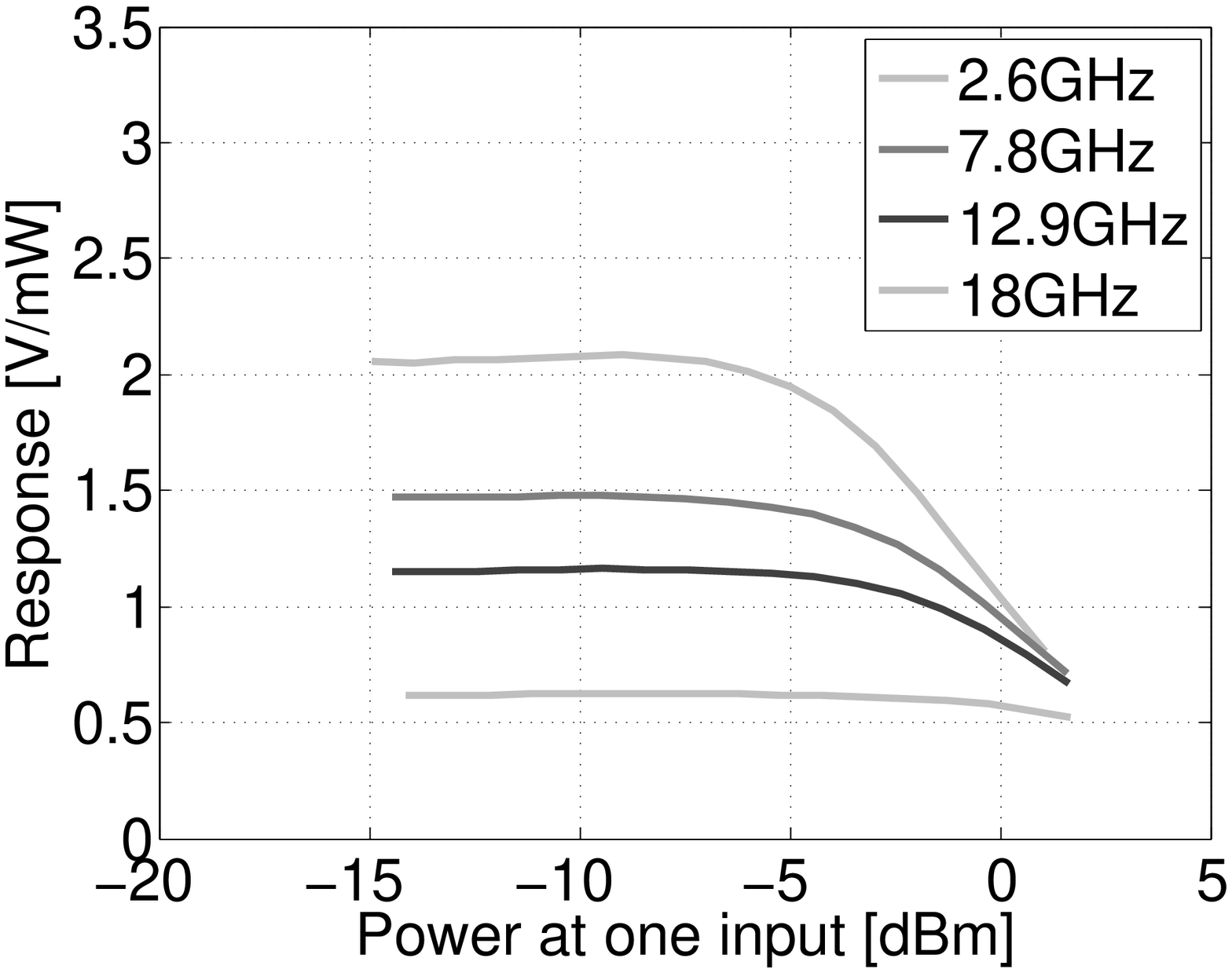}%
\\ 
\medskip
\includegraphics[width=3.04in]{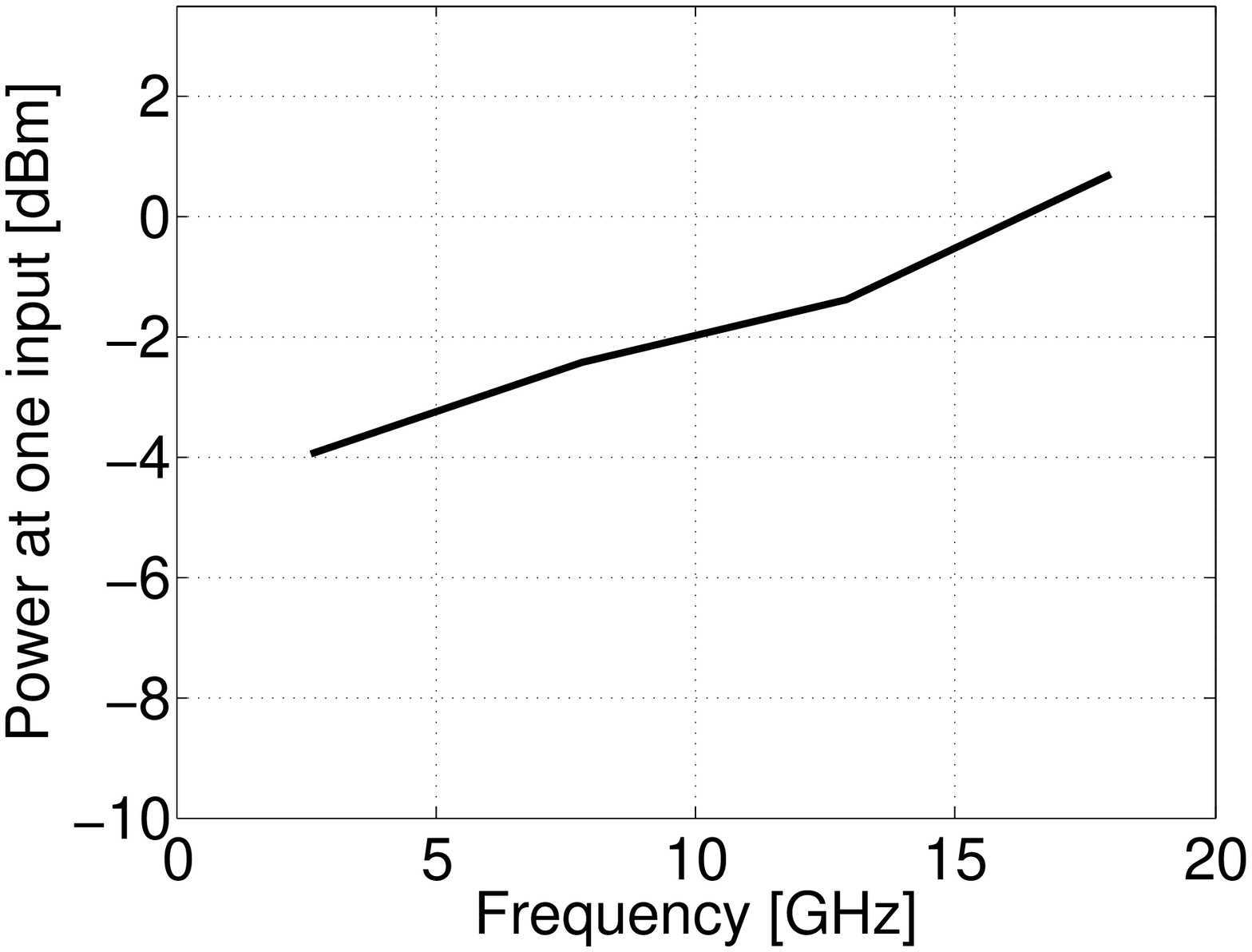}%
\end{center}
\caption{Correlator response against power for 4 different frequencies (top). Maximum CW signal power level for a compression below 10\% (bottom).}
\label{fig:power}
\end{figure}

The multiplier detects the correlated noise power, ideally producing
the true linear product of the input voltages. If the multiplier is
operating close to compression, then even small changes in the power
level will result in a change in small-signal responsivity, and hence
a change in the effective gain of the system. The power response of
the correlator is shown in Fig. \ref{fig:power}. The responsivity is
constant across the full bandwidth up to well-defined power
levels. From these data the maximum input power as a function of
frequency can be calculated. The maximum input level for 10\%
($-0.5$~dB) deviation from the linear response is shown in the lower
plot of Fig. \ref{fig:power}. In astronomical observations, where
calibration can be used to frequently re-measure the system
responsivity, a $-1$~dB compression can be used as a limit, increasing
the maximum input power level by about 2~dB. This changes the noise statistics 
by an acceptable small amount and a negligible amount of power is converted
to harmonics of the orignial signal \cite{Harris_2000}.

This defines the level and shape of a broadband input signal. For a cross-correlator, where the correlated part of the signal is deeply buried inside the uncorrelated noise from the front end, a broadband signal in the range 2--20~GHz should have a positive slope of 6~dB across the band and a signal level of approximately $-1.5$~dBm for $-0.5$~dB noise compression and 0.5~dBm for $-1$~dB noise compression. For signal to noise reasons, as explained in Section~\ref{Noise Behaviour}, the signal should not drop much below $-1.5$~dBm.

\subsection{Video Bandwidth}

The video bandwidth of the correlator is constant across the band but varies slightly between channels. A value of 10~MHz ($-3$~dB in output voltage) or 5~MHz ($-3$~dB in output power) has been measured.

\section{Noise Behaviour}
\label{Noise Behaviour}

The response of the correlator to an input signal has two components. A correlated noise-like signal of power $P_{\rm{in}}$ and spectral density $p_{\rm{in}}(\nu)$  will produce an average (DC) output voltage:

\begin{equation}
\overline{V} = \int p_{\rm{in}}(\nu)R(\nu) \rm{d} \nu,
\end{equation}

where $R(\nu)$ is the spectral responsivity of the correlator channel. The second component, the noise of the output signal, is:

\begin{equation}
V_{\rm{rms}} = \sqrt{\frac{\nu_{\rm{video}}}{\nu_{\rm{IF}}}}\cdot\int p_{\rm{in}}(\nu)R(\nu) \rm{d} \nu,
\end{equation}

where $\nu_{\rm{video}}$ and $\nu_{\rm{IF}}$  are the video and IF bandwidths, or the output and input bandwidths of the correlator; here 10~MHz and 18~GHz, respectively. The last equation is valid for correlated or uncorrelated noise and therefore  $p_{\rm{in}}(\nu)$  in (3) is the spectral power density of the correlated and uncorrelated signals added together.

\begin{figure}[h!]
\begin{center}
\includegraphics[width=3in]{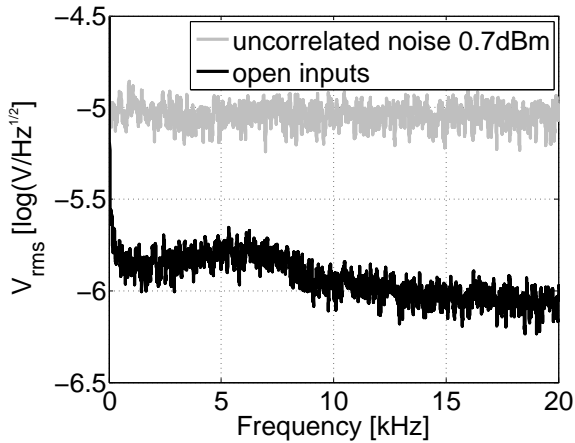}
\end{center}
\caption{Low frequency noise voltage spectrum of a typical channel with no input compared to noise response for +0.7~dBm of uncorrelated broadband (2--20~GHz) noise. The thermal noise from the signal dominates over noise introduced by the correlator itself for all frequencies of interest.}
\label{fig:noise}
\end{figure}

\begin{figure*}[t!]
\begin{center}
\includegraphics[width=6.8in]{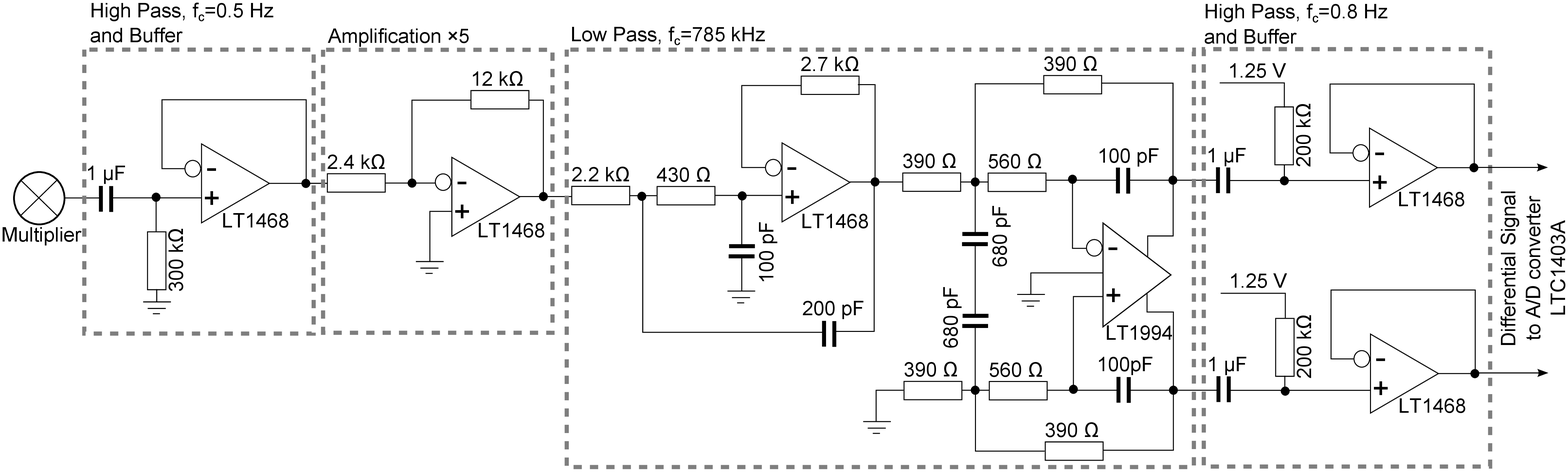}
\end{center}
\caption{
Schematic of the readout electronics of a single correlator channel. The correlator produces mainly uncorrelated noise with 25~mV RMS. Buried within is the phase-switched correlated part of the signal. In addition there is a 6~V DC output from the multipliers, which is filtered out in the first high-pass filter. Later in the signal chain a DC level of 1.25~V is added to match the 0--2.5~V input range of the A/D converters.}
\label{fig:readout}
\end{figure*}

The multipliers also produce intrinsic output noise $S_{\rm{rms}}$  without input. They are high-noise devices and some care has to be taken to find the correct signal levels and phase switching frequency in order for the overall signal-to-noise not to be dominated by noise contributions from the multipliers. Fig. \ref{fig:noise} shows the noise voltage spectrum of one channel with open inputs, along with the equivalent power spectrum when the correlator is fed with uncorrelated broadband noise at the level 0.7~dBm, simulating the (noise-dominated) signal expected from the antennas of an interferometer. All channels have very similar noise levels with a typical voltage spectral density of less than $S_{\rm{rms}}=1.5~\mu \rm{V}/\sqrt{Hz}$  at frequencies above 1--2~kHz, which is a factor of 8--10 lower in voltage (18--20~dB in power) than the thermal noise spectrum due to the signal. The $1/f$ knee frequency for the correlator noise is about 500~Hz, and the correlator noise signal does not exceed the input noise signal above 100~Hz. Therefore provided the signal is modulated by phase switching at a higher frequency than $\sim$100 Hz, the correlator will not introduce excess noise  to the measurement. 

\section{Readout Electronics}

The purpose of the readout electronics is to buffer the output signals from the multipliers, amplify and filter them before they can be digitised by a set of A/D converters. The principal signal chain after the multiplier is shown in Fig. \ref{fig:readout}. 

As described above, the multiplier output contains the phase switched correlated signal buried in uncorrelated noise. In order to have a good rejection of any unwanted correlated signals picked up after the receiver, low leakage between different orthogonal switching functions is necessary.  For this the shape of the switching function has to be sampled very accurately. This requires high sampling rates in the A/D converter (here 2.8~Msamples/sec) and careful design of the signal filters. 

To avoid aliasing, the analog electronics should filter out signals above 1.4~MHz. However, an undisturbed square wave requires high signal bandwidth and small phase distortions. Therefore a compromise between steep frequency cut-off, low phase distortions and limited circuit complexity had to be found. An active four-pole Bessel filter using a relatively small number of components was implemented.

At the lower end of the frequency band similar restrictions apply. The multiplier signal contains an offset DC voltage, which depends on the bias supply voltage and has to be filtered out. However, a high cut-off frequency produces a droop which distorts the square wave. Therefore the corner frequency of the high pass filter has to be set as low as possible, in our case below 1~Hz.

The detailed circuit layout of the readout electronics is also shown in Fig. \ref{fig:readout}. The multiplier bias voltage is 7~V and the output DC offset 6~V. The first low-pass filter which blocks this DC level consists of a capacitor of $1~\mu \rm{F}$ and a resistor of $300~\rm{k}\Omega$. This is small compared with the input impedance of the first buffer stage and therefore the high-pass 3~dB frequency is 0.5~Hz. However, the relatively high resistor value produces a small DC offset due to the input bias current of the LT1468. This is a maximum of 40~nA and therefore the offset of the first stage is about 12~mV but will be amplified in the second stage. 

In the next stage the signal is amplified slightly. For typical IF signal levels the multiplier produces a noise-like random signal of about  $V_{\rm{rms}} = 25~\rm{mV}$, and approximately 120~mV peak-to-peak. Since the input range of the ADC is 0 to 2.5~V and the highest peaks of the noise signal should be within this range, we only amplify by a factor of 5.

The next two stages make up a 4-pole active Bessel filter with a 3~dB frequency of 785~kHz and a rejection at the Nyquist frequency of 1.4~MHz of $-9.7~\rm{dB}$. The first stage is a Sallen-Key design and the second a multiple-feedback filter. In addition the second stage produces a differential signal, since this is required as input by the A/D converter.

After the filter the signal is elevated to a DC level of 1.25~V since the A/D converter’s input range is 0 to 2.5~V. Simultaneously a second high-pass filter blocks any small DC levels of the previous stages. However, the input bias current of the following LT1468 across these last resistors produces a new offset voltage. Since the signal is now differential the input offset current, the difference between the inverting and non-inverting input bias currents, sets the DC offset level. For the LT1468 this is below 50~nA. In fact the component has mainly been chosen for its small input bias current. In any case, this DC offset will be suppressed by phase-switching as long as the variations, for example due to temperature, are slower than the phase switching frequency.

\begin{figure}[h!]
\begin{center}
\includegraphics[width=3.3in]{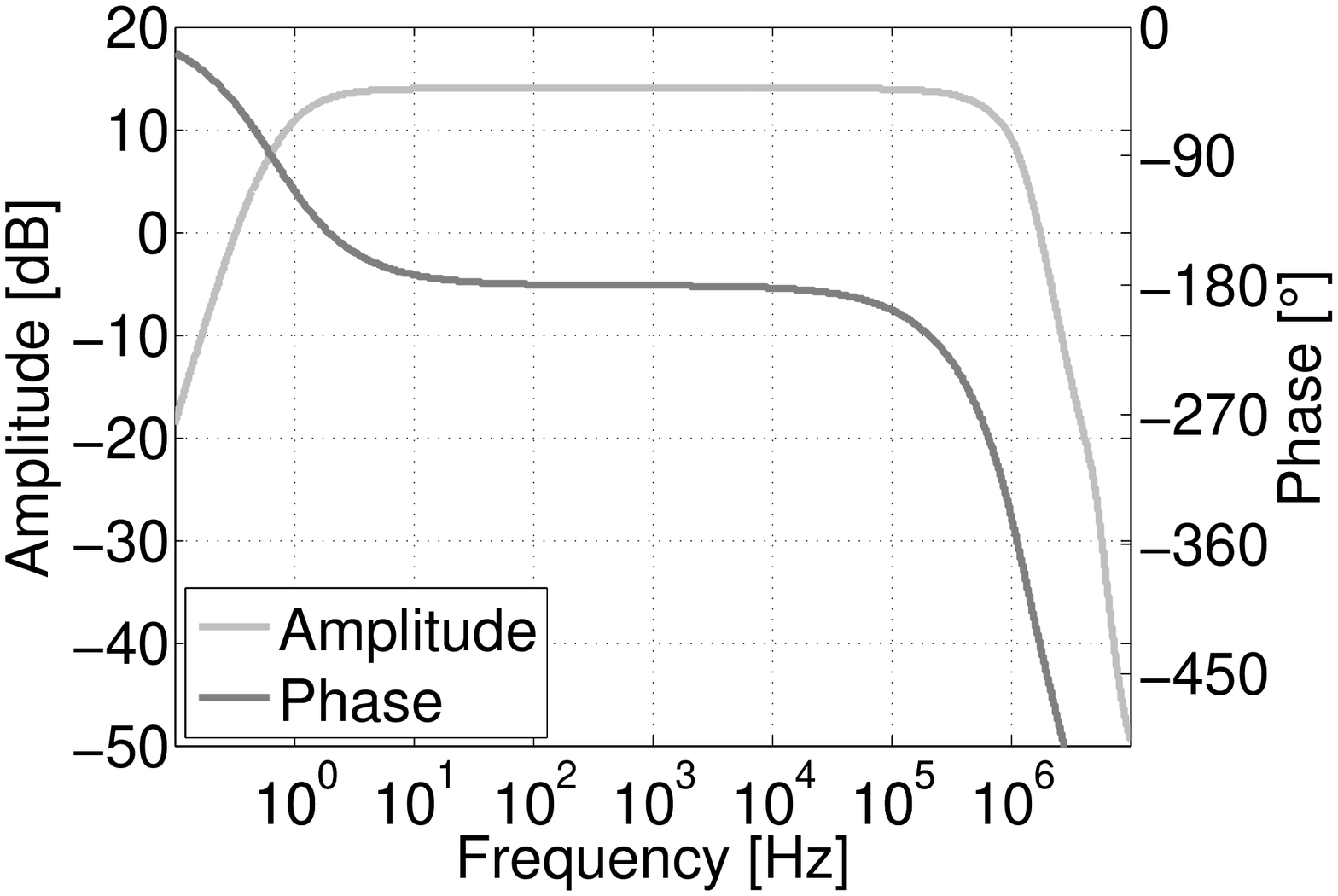}
\end{center}
\caption{
Simulation of the passband response of the analog readout circuit.}
\label{fig:simulations}
\end{figure}

The last buffer is necessary since the bias input current of the A/D converters is comparatively high and would produce a much larger voltage offset. Finally, the A/D conversion is done on a separate board using LTC1403A with 14 bit resolution. The demodulation and further integration of the signal is then done in software in an FPGA on the same board.

The simulated results of the full readout chain are shown in Fig. \ref{fig:simulations}. 
The distortion of a square wave through the full readout circuit is very small. The rise time is about $0.7~\mu \rm{s}$, which for a single baseline and 10~kHz phase switching corresponds to a reduction in correlator output of 0.7\%. The droop of the signal is only 0.05\% of the signal level after $10^{-4}~\rm{s}$. 

These values are also important for potential cross talk between different baselines since a distortion of the square wave reduces orthogonality between the phase switch functions. The effect due to rise time can be best reduced by blanking out the response of the correlator for that period. Depending on the number of baselines and therefore the period of the highest order switching function this can reduce the observing time substantially. In the case of a single baseline with 10~kHz phase switching this would reduce observing time by about 1.4\%. The droop will invariably increase cross talk between baselines since demodulating a phase switched signal with an orthogonal function does not reject the signal fully. Here it would give a rejection of about $-39~\rm{dB}$.

\section{Conclusions}

We have presented an ultra-broadband analog Fourier transform correlator. The instrument works well over 2--20~GHz with a possible maximum bandwidth of 1--23~GHz. The frequency band is split into 16 complex channels, each 1.25~GHz wide. The response of the correlator is well behaved over the band and the signal-to-noise of the system is not limited by the cross-correlator. We have also described the low frequency readout electronics necessary to digitise a fully phase-switched astronomical signal.

% use section* for acknowledgement
\section*{Acknowledgment}

The authors would like to thank J. J. John of the Oxford Physics Central Electronics group for assistance with the readout electronics and Adam Coates for assistance with the noise power spectrum measurements.

% Can use something like this to put references on a page
% by themselves when using endfloat and the captionsoff option.
\ifCLASSOPTIONcaptionsoff
  \newpage
\fi

% trigger a \newpage just before the given reference
% number - used to balance the columns on the last page
% adjust value as needed - may need to be readjusted if
% the document is modified later
%\IEEEtriggeratref{8}
% The "triggered" command can be changed if desired:
%\IEEEtriggercmd{\enlargethispage{-5in}}

% references section

% can use a bibliography generated by BibTeX as a .bbl file
% BibTeX documentation can be easily obtained at:
% http://www.ctan.org/tex-archive/biblio/bibtex/contrib/doc/
% The IEEEtran BibTeX style support page is at:
% http://www.michaelshell.org/tex/ieeetran/bibtex/
%\bibliographystyle{IEEEtran}
% argument is your BibTeX string definitions and bibliography database(s)
%\bibliography{IEEEabrv,../bib/paper}
%
% <OR> manually copy in the resultant .bbl file
% set second argument of \begin to the number of references
% (used to reserve space for the reference number labels box)
%\begin{thebibliography}
\bibliographystyle{IEEEtran}
% argument is your BibTeX string definitions and bibliography database(s)
\bibliography{IEEEabrv,gubbins}

%\bibitem{IEEEhowto:kopka}
%H.~Kopka and P.~W. Daly, \emph{A Guide to \LaTeX}, 3rd~ed.\hskip 1em plus
%0.5em minus 0.4em\relax Harlow, England: Addison-Wesley, 1999.

%\end{thebibliography}

% biography section
% 
% If you have an EPS/PDF photo (graphicx package needed) extra braces are
% needed around the contents of the optional argument to biography to prevent
% the LaTeX parser from getting confused when it sees the complicated
% \includegraphics command within an optional argument. (You could create
% your own custom macro containing the \includegraphics command to make things
% simpler here.)
%\begin{biography}[{\includegraphics[width=1in,height=1.25in,clip,keepaspectratio]{mshell}}]{Michael Shell}
% or if you just want to reserve a space for a photo:

\begin{IEEEbiographynophoto}{Christian M. Holler}
was born in southern Germany in 1973. He studied physics at the Ludwig-Maxmilians-Universit\"{a}t in Munich, Germany and earned his Diplom in 1999. In 2003 he received his PhD in the field of astrophysics and instrumentation at the Cavendish Laboratories, Cambridge University, United Kingdom. After returning to Munich, Germany he successfully pursued entrepreneurial work. In early 2007 he joined the experimental cosmology group at the University of Oxford as a postdoctoral research assistant, where he was involved in several instrumentation projects. In September 2009 he was appointed a Professor at the University of Applied Sciences in Esslingen, Germany. 
\end{IEEEbiographynophoto}
\begin{IEEEbiographynophoto}{Michael E. Jones}
was born in Wrexham, UK, in 1966 and studied Natural Science and Electrical Science at Cambridge University. He obtained a PhD in Radio Astronomy at the Cavendish Laboratory, Cambridge in 1990 and subsequently worked there as a Research Associate and Senior Research Associate. In 2005 he was appointed first as a Lecturer and then as Professor of Experimental Cosmology at the University of Oxford Department of Physics. 
\end{IEEEbiographynophoto}
\begin{IEEEbiographynophoto}{Angela C. Taylor}
was born in Dartford, UK in 1975 and studied Natural Sciences at Cambridge University, earning a PhD in astrophysics at the Cavendish Laboratory Cambridge in 2002. She continued to work in the field of astrophysics and instrumentation as a postdoctoral research assistant at the Cavendish Laboratory until 2005 when she moved to the University of Oxford as a founding member of the experimental cosmology group. Since January 2005 she has been an STFC Postdoctoral Research Fellow and then a Royal Society Dorothy Hodgkin Research Fellow working on a range of experimental cosmology projects at the University of Oxford.
\end{IEEEbiographynophoto}
\begin{IEEEbiographynophoto}{Andrew I. Harris}
received his Ph.D. in Physics from the University of California at Berkeley in 1986.  He is a Professor in the Department of Astronomy and an Affiliate Professor in the Department of Electrical and Computer Engineering at the University of Maryland, where he specializes in  experimental astrophysics and astronomical instrumentation.
\end{IEEEbiographynophoto}
\begin{IEEEbiographynophoto}{Stephen A. Maas}
earned his Ph.D. in Electrical Engineering from the University of California at Los Angeles in 1984.  Since then, he has been involved in research, design, and development of low-noise and nonlinear microwave circuits and systems at the National Radio Astronomy Observatory (where he designed the receivers for the Very Large Array), Hughes Aircraft Co., TRW, and the Aerospace Corp. For several years he was a member of the Electrical Engineering Faculty at UCLA. He is now Director of Technology at AWR, specializing in nonlinear circuit simulation technology. He teaches periodically at UCLA, both regular and extension courses.  He is the author of several books on mixers and other nonlinear devices.
\end{IEEEbiographynophoto}

% You can push biographies down or up by placing
% a \vfill before or after them. The appropriate
% use of \vfill depends on what kind of text is
% on the last page and whether or not the columns
% are being equalized.

%\vfill

% Can be used to pull up biographies so that the bottom of the last one
% is flush with the other column.
%\enlargethispage{-5in}

% that's all folks
\end{document}